\documentstyle[aps,prl,multicol]{revtex}
\DeclareSymbolFont{AMSb}{U}{msb}{m}{n}
\DeclareSymbolFontAlphabet{\mathbb}{AMSb}
\newcommand{\complex}{\kern.1em{\raise.47ex\hbox{
            $\scriptscriptstyle |$}}\kern-.40em{\rm C}}
\newcommand{\ket}[1]{\left\vert #1 \right\rangle}
\newcommand{\kernel}[3]{\left\langle #1\left\vert #2\right\vert#3 \right\rangle}

\newcommand{\cmp}{Comm.\ Math.\ Phys.}

\newcommand{\abs}[1]{\left\vert #1\right\vert}

\input epsf

\title{ Hofstadter butterfly as Quantum phase diagram}
\begin{document}
\author{D. Osadchy and J.~E.~Avron}
\address{Department of Physics, Technion, 32000 Haifa, Israel}
\maketitle
\begin{abstract}
The Hofstadter butterfly is viewed as  a quantum phase diagram
with infinitely many phases, labeled by their (integer) Hall
conductance, and a fractal structure. We describe various
properties of this phase diagram: We establish Gibbs phase rules;
count the number of components of each phase, and characterize the
set of multiple phase coexistence.
\end{abstract}

\pacs {PACS numbers: 73.43.-f, 73.43.Nq}
\begin{multicols}{2}

{\em Introduction.---} Azbel \cite{azbel} recognized that the
spectral properties of two-dimensional, periodic, quantum systems
have sensitive dependence on the magnetic flux through a unit
cell. A simple model conceived by Peierls and put to the
eponymous Harper as a thesis problem,
gained popularity with D. Hofstadter Ph.D. thesis
\cite{hofstadter}, where a wonderful diagram, reminiscent of a
fractal butterfly, provided a source of inspiration and  a tool
for spectral analysis
\cite{bellissard-gl,bs,helffer,last,thouless,wiegmann,wilkinson}.

The Hofstadter butterfly can also be viewed as the quantum (zero
temperature) phase diagram for the integer quantum Hall effect. It
is a fractal phase diagram with infinitely many phases
\cite{as,ketzmerick}. The diagram leads to certain natural
questions: Count the number of components of a given phase;
Classify which phases coexist and where. It also leads to the
general question: What form does the Gibbs phase rule
\cite{israel,ruelle} take for quantum phase transitions.

Fractal phase diagrams and/or infinitely many phases appear in
dynamical systems \cite{bak,ott}. In classical lattice systems
fractal phase diagrams \cite{israel,griffiths,parisi,ruelle} are
commonly viewed as a pathology due to either long range
interactions, or, as is the case for spin glasses,  loss of
translation invariance. The Hofstadter model, when viewed as a
statistical mechanical model, is both short range and translation
invariant in a natural way. But, it is quantum and the translation
group is non-commutative. It suggests that fractal phase diagram
may be more common in quantum phase transitions than in classical
phase transitions.

{\em The Hofstadter Model.---} The model conceived by Peierls has
two versions. For the sake of concreteness we shall focus here on
the tight binding version. On the lattice ${\mathbb{Z}}^2$, define
magnetic shifts
\begin{eqnarray}\label{shifts}
& (U\psi)(n,m)=\psi(n-1,m) \nonumber \\
& \big(V(\Phi)\psi\big)(n,m)=e^{2\pi i\,n\,\Phi}\,\psi(n,m-1) \\
&  n,m\in \mathbb{Z} \nonumber
\end{eqnarray}
$2\pi\Phi$ is the magnetic flux through a unit cell.
The
Hofstadter model is
\begin{equation}\label{harper}
  H(\Phi,a,b)=a(U+U^*)+b\big(V(\Phi)+V^*(\Phi)\big),
\end{equation}
where $a,b>0$ are ``hopping'' amplitudes. $a=b$ is called the
self-dual case \cite{aubry} and we shall focus on that case below.
We set $H(\Phi)=H(\Phi,1,1)$.

$U$ and $V$, and therefore also $H(\Phi)$, commute with the (dual)
magnetic translations ${\cal U}$ and ${\cal V}$,
\begin{eqnarray}\label{magnetic}
& ({\cal U}\psi)(n,m)=\psi(n,m-1) \nonumber \\
& ({\cal V}\psi)(n,m)= e^{2\pi i\Phi m}\,\psi(n-1,m)
\end{eqnarray}
This makes $H(\Phi)$ translation invariant in a natural way. The
group of magnetic translations \cite{zak} is non-commutative:
\begin{equation}\label{connes}
  U^*V^*UV={\cal U VU^*V^*}= e^{-2\pi i\Phi}.
\end{equation}
The one particle representation of the Hofstadter model,
Eq.~(\ref{harper}), is natural for spectral studies. In the
context of statistical mechanics the second quantized
representation of the model is also instructive because it makes
it clear that the model has short range, in fact, only on site and
nearest neighbors, interactions. The fractal features  of the
phase diagram are, therefore, not a consequence of long range
forces, as in some classical statistical mechanics models. The
second quantized form is:
 \begin{equation}\label{sq}
{\cal H}(\Phi,\mu)= \sum e^{i\gamma(nm;n'm')}
a^\dagger_{nm}a^{\phantom{\dagger}}_{n'm'} +\mu\sum
a^\dagger_{nm}a^{\phantom{\dagger}}_{nm}
\end{equation}
where
\begin{equation}\label{gamma}
  e^{i\gamma(mn;m'n')}=
  \cases{
    1 & $ {n-n'}=\pm 1,\ m=m'$; \cr
    e^{\pm 2\pi i n\Phi} &  $ {m-m'}=\pm 1,\ n=n'$; \cr
    0 & otherwise.
  }
\end{equation}
$\mu$ is the chemical potential and $a^\dagger, a$ are the usual
Fermionic operators.

Let us recall few elementary features of the spectrum:
\begin{eqnarray}\label{spectral}
 S\big(H(\Phi)\big)&=&- S\big(H(\Phi)\big)=-
 S\big(H(1-\Phi)\big).
\end{eqnarray}
 The first is a consequence of ${\mathbb{Z}}^2$ being
bipartite, and the second is a consequence of time reversal.
Together, they imply a four-fold symmetry, manifest in the
Hofstadter butterfly.

The electronic density, $\rho(\Phi,\mu)$, (=integrated density of
state), is
\begin{equation}\label{dos}
\rho(\Phi,\mu)= \kernel{0}{\theta\big(\mu-H(\Phi)\big)}{0}
\end{equation}
where $\ket{0} $  is Kroneker delta at the origin.
$0\le\rho(\Phi,\mu)\le 1$, is an increasing function of $\mu$.
$\theta$ is the usual step function.

The gaps in the spectrum are labeled by an integer, $k$, which is
a solution of \cite{thoulessBOOK,bellissard-gl},
\begin{equation}\label{gap-labeling}
\Phi\,  k=\rho\ mod \ 1.
\end{equation}
 $k$ is the Hall conductance. We picked the letter $k$ because it
is naturally associated with an integer, and it is also the first
letter in von Klitzing name. By Eq.(\ref{gap-labeling}) and
Eq.~(\ref{spectral}) :
\begin{equation}\label{symmetry}
k(\mu,\Phi)=-k(\mu,1-\Phi)=-k(-\mu,\Phi).
\end{equation}
which implies a four fold (anti) symmetry of the butterfly.

We shall assume that the Ten Martini conjecture
\cite{jitomirskaya} holds. Namely, that for all irrational
$\Phi$'s, all the gaps are open, so Eq. (\ref{gap-labeling}) has
$\rho$ in an open gap for all $k \in \mathbb{Z}$.

\begin{figure}[h]
\center{ \epsfxsize=3.2in \epsfbox{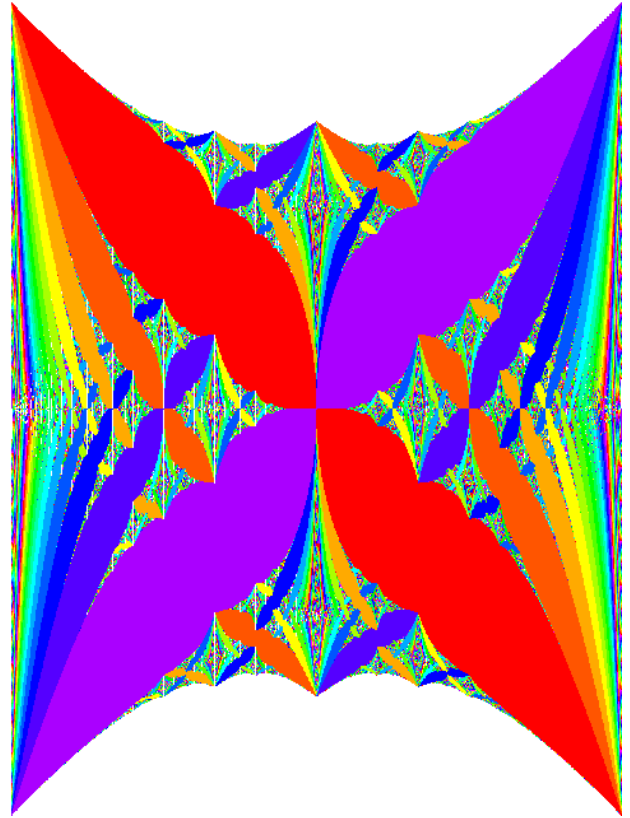} }
\caption{Hofstadter colored butterfly} \label{color}
\end{figure}

Fig. \ref{color} shows the Hofstadter butterfly, color coded
according to the Hall conductance. The color coding counts the
Hall conductance modulo 16. Zero Hall conductance and the spectrum
are left blank. The gross features of the diagram are associated
with small integers where the color coding is faithful.

The colored picture emphasizes the gaps while the standard
Hofstadter butterfly  emphasis the spectrum. The colored figure is
prettier and  displays the regular aspects of the diagram: Gaps
are better behaved than spectra. The colored diagram is also more
faithful  to certain spectral characteristics. For example, the
spectrum is a small set, (in fact, one of zero Lebesgue measure),
something that is manifest in the colored diagram, but is less
obvious from the usual Hofstadter butterfly which plots the
spectrum.

We also  broke with tradition in that the colored Hofstadter
butterfly is rotated by $90^\circ$: In Fig. \ref{color} the
horizontal axis is $\Phi$ and the vertical axis is the energy, or
$S\big(H(\Phi)\big)$. The reason we chose to do so is that this
way emphasizes the fact that phase boundaries are functions (of
$\Phi$).

We denote by $P(k)$ the  $k$-th phase. Formally,

\begin{equation}\label{set}
P(k) = \left\{ \Phi, \mu \big\vert \ \Phi k =\rho(\Phi,\mu)\ mod\
1 , \mu\notin S\big(H(\Phi)\big) \right\}
\end{equation}
$P(k)$ is an open set in the $(\Phi,\mu)$ plane, with a finite
number of components. For example, $P(1)$ is two of the four big
wings of the butterfly. We call $P(k)$ a {\em pure phase} and
denote its number of components $|P(k)|$. The closure of the pure
phase is denoted $\bar P(k)$ and the phase boundary is $\partial
P(k)$.  We call $\cup_k\,\partial P(k)$ the total boundary.

{\em Counting components.---} The $k$-th pure phase is made of
several components. The $k=0$ phase (blank) has two components.
For $k\neq 0$ the number of components is
\begin{equation}\label{counting}
\abs{P(k)}=\sum^{2\abs{k}}_{j=1}\phi(j)= 12\, \frac
 {k^2}{\pi^2}+O\left(k \log k\right)
\end{equation}
where $\phi(j)$ is Euler (totient) function. Recall that $\phi(j)$
counts the number of integers, up to $j$ (and including 1), that
are prime to $j$: {$\phi(1)=1,\ \phi(2)=1,\ \phi(3)=2$ etc.}.

To prove Eq.~(\ref{counting}) note first that  from
Eq.~(\ref{gap-labeling})
\begin{equation}\label{unique}
k(\rho,\Phi)=k(\rho ',\Phi)\Rightarrow\rho=\rho'.
\end{equation}
Hence, a given color would appear {\em at most} once on any
vertical line of fixed $\Phi$.

 When $\Phi=\frac p q$ with $gcd(p,q)=1$ ($gcd(p,q)$ is the greatest
common divisor of $p$ and $q$), the spectrum of the Hofstadter
model has $q$ finite bands and $q-1$ gaps which are all open
intervals (except when $q$ is even the central gap is closed). We
then number the gaps by their natural order $1,\dots,q-1$. The
semi-infinite interval below the spectrum is, formally, the 0-th
gap, and the semi-infinite interval above the spectrum as the
$q$-th gap.
 Eq.~(\ref{gap-labeling}), for the j-th
gap, can be written as
\begin{equation}\label{diophantine2}
p\, k= j\ mod\ q.
\end{equation}
Given $p,q$ and $j$ the equation has a unique solution for each
open gap such that $\abs{k}<\frac q 2$. In particular, once $q$
has been fixed, the Hall conductance takes all (non-zero) integer
values, from $-\lfloor\frac q 2\rfloor$ to $\lfloor\frac q
2\rfloor$, and each value appears once. For even $q$, the central
gap is closed and formally can be assigned a value of $\pm \frac q
2$.

The number of components of the $k$-th phase is the same as the
number of flux values, $\Phi_\ell$, which accommodate the
wing-tips, minus one.
 The $k$-th wings-tips are located at those values of $\Phi$
where the $k$-th color is absent (the $k$-th color {\em is}
present in any small neighborhood of those values of $\Phi$,
because this neighborhood contains fractions with arbitrary large
$q$'s). Given $k\neq 0$, it must appear once on a horizontal
interval with $\Phi=\frac p q$ provided $\frac q 2>\abs{k}$. The
tips of a wing with a given color must, therefore, be located at
those values of $\Phi$ which do not admit $k$ as solution of
Eq.~(\ref{diophantine2}) for any $\rho$. In other words, the
wing-tips lie at those values of $\Phi$ where $q$ is too small to
accommodate $\abs{k}$. This is the finite set, a Farey sequence,
\begin{eqnarray}\label{osadchy}
F_{2\abs{k}} & = & \left\{\frac p q\ \Big\vert\ 0 \le p \le q, \,
gcd(p,q)=1,\, q\le 2\abs{k}\right\} \\ & = &
\bigcup_{q=1}^{2\abs{k}} \left\{\frac p q\ \Big\vert\ 0 < p\le q,
\, gcd(p,q)=1\right\} \bigcup \left\{ 0 \right\} \nonumber
\end{eqnarray}
Let $|F|$ be the number of elements in $F$. Then from
(\ref{osadchy}),
\begin{equation}\label{recursion}
|P(k)| = |F_{2\abs{k}}| - 1 = \sum_{q=1}^{2\abs{k}} \phi(q)
\end{equation}
essentially from the definition of the Euler function.

  The asymptotic
expansion for the sum in Eq.~(\ref{counting}) is taken from
\cite{totient}.

{\em Pure phases and Phase boundaries.---}  In thermodynamics and
statistical mechanics,  Gibbs phase rule is a statement about the
structure of pure phases and their boundaries. A weak form of the
Gibbs phase rule says that pure phases are a set of full measure;
two phases coexist, generically, on a set of Hausdorff
co-dimension one etc. \cite{israel}. This is the form that one
gets if one considers general convex functions for thermodynamic
potentials. The number of coexisting phases is related to the
dimension of tangent planes, and the Gibbs phase rule is a
consequence of theorems about convex functions. There is a
stronger form of the rule \cite{ruelle} which posits, in addition,
that the sets are (locally) manifolds. This form is a consequence
of additional regularity of the thermodynamic potentials.

Gibbs phase rule is a consequence of the convexity of
thermodynamic potentials and so is ultimately based on the second
law of thermodynamics. It has nothing to say about the zero
temperature phase diagram of quantum phase transitions in general,
\cite{ss}, and the Hofstadter model in particular (because the
entropy vanishes identically). A question that arises is then what
form might Gibbs phase rule take for quantum phase transition. The
Hofstadter model restricts what could and what could not be true
in general. As we shall see, the phase diagram of the Hofstadter
model turns out to satisfy only a weak form of the Gibbs phase
rules.

Fig. \ref{color} suggests that the set of unique phase is a set of
full measure and that the phase boundaries, though fractal, are
not too wild.  More precisely, we have:

{\bf Gibbs-like phase rule:} {\em   The phase diagram of the
self-dual Hofstadter model is such that pure phases, labeled by
their Hall conductances, are full measure; phase boundaries are
{\em not} manifolds---they are nowhere differentiable-- but they
are almost so in the sense that their Hausdorff co-dimension is
integral, in fact:
\begin{equation}\label{conjecture}
  dim_H\big(\partial P(k)\big) =1.
\end{equation}
Since the number of phases is countable  the total phase boundary
${\displaystyle \cup_m\, \big(\partial P(m)\big)}$, is a set of
Hausdorff dimension one as well. Finally, infinitely many phases
coexist on a countable set, and therefore a set of Hausdorff
dimension zero. }

The first part of the Gibbs-like phase rule is an easy consequence
of a result of Last \cite{last}, which states that
$\abs{S\big(H(\Phi)\big)}=0$ for a set of $\Phi$ of full measure.
That phase boundaries are nowhere differentiable follows from
results of Wilkinson \cite{wilkinson}, Rammal, and Helffer and
Sj\"ostrand \cite{helffer} who showed that the phase boundaries
$\partial P(k)$ possess distinct left and right tangent at every
rational $\Phi$. That the Hausdorff dimension of the boundary is
one follows from results of Bellissard \cite{bellissard} who
showed that away from the wings tips, $\partial P(k)$ can be
represented by a functions of $\Phi$ that are {\em uniformly}
Lipshitz. By standard results, \cite{falconer} it then follows
that the Hausdorff dimension is one. The set of infinite phase
coexistence is analyzed below.

{\em Coexistence.---} In \cite{ott} the term lakes of Wada was
used to describe dynamical systems with the property that any
point on the boundary of the one basin of attraction is also on
the boundary of all other basins. We shall say that a system is
almost Wada of order $m$ if every circle that contains two pure
phases contains $m$ pure phases.

The Hofstadter butterfly is almost Wada of infinite order. This is
seen from the figure, and can also be shown to follow from
Eq.~(\ref{gap-labeling}).

We say that the two pure phases, $P(m),\ P(n)$, coexist on
\begin{equation}\label{co}
  C(m,n)= \partial P(m)\cap \partial P(n)
\end{equation}
No two phase coexist for any irrational flux. This  is easily seen
from Eq.~(\ref{gap-labeling}): For irrational $\Phi$ the electron
density $\rho$ takes a dense set of values in the gaps. Therefore,
any two phases, $P(m)$ and $P(n)$ are separated by infinitely many
other phases. It follows that the set of phase coexistence is a
countable set, and so of zero Hausdorff dimension.

The following result gives a complete characterization of phase
coexistence:

0 {\bf Proposition:} {\em
Consider a point $x\in\partial P(k)$ with $\Phi(x)=\frac p q$ with
$gcd(p,q)=1$. Then $x \in\partial P(k+\ell q)$ for all $\ell \in
\mathbb{Z}$. Moreover $x \notin \partial P(k')$ if $k' \neq k+\ell
q $ for each $\ell \in \mathbb{Z}$}

{\bf Proof:} Since $gcd(p,q)=1$ the equation
\begin{equation}
  pa-qb=1
\end{equation}
has a solution with integer $a$ and $b$ (where $a$ is non-unique
mod $q$). Let $\frac {p_n}{q_n}= \frac{np-b}{nq-a}$ with
$n\in\mathbb{Z}$. Then
\begin{equation}
  p_n\, q-q_n\, p=1.
\end{equation}
From Eq.~(\ref{diophantine2}) it follows that each band at $\frac
{p_n}{q_n}$ carries Hall conductance $q$ {\tt(mod $q_n$)}.

Now consider a point $x$ on the right boundary of $P(k)$. We shall
first show that $x$ is also on the left boundary of $P(k+q)$.
Since $P(k+q)$ has a finite set of wing-tips, Eq.~(\ref{osadchy}),
when $n$ is large enough, the gap with label (=Hall conductance)
$k+q$ at flux $\frac {p_n}{q_n}$ must be open, and must remain
open for all large $n$. By a bound of Last and Wilkinson \cite{lw}
for  the total width of the spectrum at $\frac {p_n}{q_n}$, each
band is small and hence
\begin{equation}\label{wl}
dist(P(k),P(k+q))< \frac {24}{q_n}.
\end{equation}
Taking $n\to\infty$ we see that $x\in\partial P(k+q)$ as claimed.

By considering the next band we shall now show that $x$ also lies
on the boundary of $P(k+2q)$. Now, $P(k+2q)$ at  $\frac
{p_n}{q_n}$  is separated from $P(k)$ by two bands and a gap. The
bands are small by the Last Wilkinson bound. The gap is also small
by the H\"older continuity of the spectrum:
\begin{equation}\label{gap}
  \abs{gap} < 18\sqrt{\frac {p_n}{q_n}-\frac p q}=\frac {18} {\sqrt{q
  q_n}}
\end{equation}
and from this
\begin{equation}
dist(P(k),P(k+2q))< \frac {24}{q_n}+\frac {18}{\sqrt{q q_n} }.
\end{equation}
Taking the limit $n\to\infty$ yield the result. The argument can
be repeated for any $P(k+\ell q)$ with $\ell$ finite and positive.
Negative values $P(k-\ell q)$ are obtained by letting $n\to
-\infty$ in the argument above.

For the left boundary point of $P(k)$, $n\to\infty$ will give
$P(k-\ell q)$ and $n\to-\infty$ will give $P(k+\ell q)$. This
formula applies also to the phase $k=0$, with it's right boundary
being the leftmost point of the spectrum and vice-versa. The Hall
conductance for the middle gap with even $q$ (which is closed) is
formally $\pm \frac q 2$, so it is common to phases $P(\pm \frac
q 2 +\ell' q)$ which is the same as $P(\frac q 2 +\ell q)$ for
$\ell,\ell' \in \mathbb{Z}$.

The second part of the proposition follows from the equality
\begin{displaymath}
\bigcup_{k\le|\frac q 2|} \bigcup_{\ell \in \mathbb{Z}} P(k+\ell
q)=\bigcup_{k \in \mathbb{Z}} P(k)
\end{displaymath}
and the fact the for fixed $\Phi$, each Hall conductance $k$ can
appear only once.

We thank M.V. Berry, S. Jitomirskaya, A. Kamenev, H. Kunz, D.
Hermann, B. Simon, H. Schultz-Baldes, Y. Last, R. Ketzmerick and
A. van Enter for useful discussions. This research was supported
in part by the Israel Science Foundation, the Fund for Promotion
of Research at the Technion.

\end{multicols}

\end{document}